\documentclass[prl,aps,preprint,bibtex,showpacs]{revtex4}
\usepackage{graphics}

\begin{document}
\title{Rotation in string cosmology}

\author{Yuri N.\ Obukhov\footnote{On leave from: Department
of Theoretical Physics, Moscow State University, 117234 Moscow,
Russia}, Thoralf Chrobok, and Mike Scherfner\footnote{Institut f\"ur
Mathematik, Technische Universit\"at Berlin, Str. d. 17. Juni 136,
D-10623 Berlin}}
\affiliation {Institut f\"ur Theoretische Physik, Technische
Universit\"at Berlin, Hardenbergstr. 36, D-10623 Berlin, Germany}

\begin{abstract}
We describe exact cosmological solutions with rotation and expansion
in the low-energy effective string theory. These models are spatially
homogeneous (closed Bianchi type IX) and they belong to the family of
shear-free metrics which are causal (no closed timelike curves are
allowed), admit no parallax effects and do not disturb the isotropy
of the background radiation. The dilaton and the axion fields are
nontrivial, in general, and we consider both cases with and without
the central charge (effective cosmological constant).
\end{abstract}
\pacs{PACS no.: 04.20.Cv; 04.20.Jb; 98.80.Cq; 98.80.Hw} 


\maketitle

\section{Introduction}

The study of cosmological models in the framework of the low-energy
effective string theory attracts considerable attention now, see, e.g.,
\cite{gas,tseytlin,nonc,cop,bar2,eas,bat,bar3,bar1,kanti,chen1,chen2}.
The (super)string theory provides a scheme for the unification of all
the physical interactions, including gravity. In the perturbative
approach to the full quantum string theory, one derives the effective
action for the metric $g$, scalar dilaton field $\phi$ and the 3-form
axion field $H$ which represents the massless field background. The
lowest order effective string model describes the Einstein gravity
coupled to $\phi$ and $H$.

The string cosmology is of interest, primarily, in connection with
the so called pre-big bang phase of the universe's evolution
\cite{gas,tseytlin}, and there are certain expectations that it can
improve our understanding of the physics in the range of energies
between the string scale and the grand unification (GUT) scale.

Recently the cosmological string models with the nontrivial global
rotation have been investigated in \cite{bar1,kanti,chen1}. In
particular, in \cite{bar1} the attention was confined to the
stationary G\"odel type models without expansion, and it was
demonstrated that the causal solutions (without closed timelike
curves) exist in the first order (in the string tension $\alpha'$)
effective models. These results were subsequently generalized in
\cite{kanti} to the ``charged'' solutions for the models with scalar
and electromagnetic fields included. In an attempt to develop a more
realistic cosmological model, the possibility of a rotating {\it and}
expanding universe was investigated in \cite{chen1} within the
framework of the perturbation theory. A further logical step which
can obviously bring a further insight into the rotating cosmologies,
should be thus a study of the existence of {\it exact} solutions with
rotation and expansion. Here we describe such exact models in the
low-energy effective string theory, demonstrating that the absence of
the closed timelike curves can be combined with the nontrivial
expansion and rotation of the universe.

More specifically, in our study we consider the homogeneous Bianchi
type IX models with rotation and expansion. The homogeneous Bianchi
models in string cosmology were investigated rather extensively in
\cite{cop,bar2,bat,bar3,chen2}. However, the main efforts were
concentrated on the analysis of the evolution of the anisotropy
described by the nontrivial shear. For example, the study \cite{bar3}
of the anisotropic Bianchi type has demonstrated the absence of the
chaotic behavior of the solutions in an anisotropic string cosmology
with shear. In contrast, our models are {\it shear-free}, and thus we
can look into the pure rotational effects in cosmology.

It seems worthwhile to mention that the study of the shear-free
models is of interest in itself within the framework of mathematical
cosmology. In particular, it is rather difficult to construct
physically viable models with nontrivial rotation and expansion in
absence of shear: In many cases the traditional sources like an ideal
fluid, electromagnetic and scalar fields, etc., simply do not allow
for such solutions. At best, it is possible to have either expansion
without rotation, or a stationary rotating model without expansion.
This constitutes the essence of the so called shear-free conjecture,
see an overview in \cite{seno}. It is thus highly interesting to
discover that the string cosmology (with the standard sources
described by the dilaton $\phi$ and the axion $H$) does admit exact
shear-free solutions with rotation and expansion.

The new solutions are completely causal, in the sense that the closed
timelike curves are absent in these rotating models. In this way, we
obtain an important extension of the earlier results
\cite{bar1,kanti,chen1} by demonstrating the existence of an exact
(not a perturbation) rotating, expanding, and causal universe in
string cosmology.

Another aspect of our results deserves a special attention. Our
models describe a spatially {\it closed} universe, and we are thus
returning to the old problem of the existence of the truly
anti-Machian world models. The early model of Ozsv\'ath and
Sch\"ucking \cite{osch} provides an example of such anti-machian
cosmology belonging to the Bianchi type IX. Obviously, the model of
Ozsv\'ath and Sch\"ucking was to a great extent of academic interest
only because it described a nonexpanding world. Moreover, it has been
subsequently demonstrated that one can reveal a certain compensation
mechanism which yields the total angular momentum equal zero for such
a closed world \cite{king,nine}, thus effectively removing its status
of the truly anti-Machian cosmology. Our new solution describes a
closed {\it and} expanding world with rotation, thus giving a
qualitatively new example of an anti-Machian model, cf. also \cite{infl}.
Furthermore, as far as we can see, there is no compensation scheme
similar to \cite{king,nine} for the closed rotating models constructed
in the present paper.

\section{String cosmology}

The lower-energy effective string theory action reads, in the compact
notation of exterior calculus,
\begin{equation}
S =\int e^{-2\phi}\left(\widehat{R}\,\widehat{\eta} + 4\,d\phi\wedge
{}^{\widehat{\star}} d\phi - {\frac 12}\,H\wedge{}^{\widehat{\star}}
H + 2\Lambda\,\widehat{\eta}\right).\label{Lstr}
\end{equation}
Here $\widehat{\eta} = {}^{\widehat{\star}}1$ is the $n$-form of the
spacetime volume, $\phi$ is the scalar dilaton field, and the 3-form
field $H=dB$ with the 2-form potential $B$ is the axion. The quantities
with hats refer to the so-called string frame of the 1-form cobasis
$\widehat{\vartheta}^\alpha$. The effective cosmological constant
(or the central charge) is related to the critical dimension of the
(super)string theory: $\Lambda = (26 - D)/3\alpha'$ for boson string
and $\Lambda = (15 - 3D/2)/3\alpha'$ for heterotic or superstring.
Although this quantity vanishes in critical dimension, it may be
nontrivial when a string couples to a conformal theory, for example.
Non-rotating solutions in non-critical string cosmologies have been
extensively studied in \cite{tseytlin,nonc,eas}. Like in the latter
references, we thus consider $\Lambda$ as an arbitrary parameter.
Finally, we will assume that the original higher dimensional theory
is reduced to $D= 4$ by means of a compactification mechanism.

The field equations read, in components
\begin{eqnarray}
4\,\widehat{D}_\alpha\widehat{D}_\beta\phi +
2\,\widehat{R}_{\alpha\beta} - {\frac 1 2}\,H_{\alpha\mu\nu}
H_\beta{}^{\mu\nu} &= &0,\label{einS}\\
4\,\widehat{D}_\mu\widehat{D}^\mu\phi - 4\,(\widehat{D}_\mu\phi)
(\widehat{D}^\mu\phi) + \widehat{R} - {\frac 1 {12}}
\,H_{\alpha\mu\nu}H^{\alpha\mu\nu} + 2\Lambda &= &0,\label{dilS}\\
d\,{}^{\widehat{\star}} H - 2\,d\phi\wedge{}^{\widehat{\star}} H
&= &0.\label{axionS}
\end{eqnarray}

The conformal transformation $\widehat{\vartheta}^\alpha =  e^{\phi}
\,\vartheta^\alpha$ from the string frame to the Einstein frame
brings the action (\ref{Lstr}) to
\begin{equation}\label{LstringE}
S =  \int\left(R\,\eta - 2\,d\phi\wedge{}^\star d\phi - {\frac 12}
\,e^{-4\phi}\,H\wedge{}^\star H + 2\Lambda\,e^{2\phi}\,\eta\right).
\end{equation}
The corresponding field equations read
\begin{eqnarray}
R_{\alpha\beta} - {\frac 1 2}\,R\,g_{\alpha\beta} - \Lambda
\,e^{2\phi}\,g_{\alpha\beta}  &= & T_{\alpha\beta},\label{einE}\\
D_\mu D^\mu\phi + {\frac 1 {12}}\,e^{-4\phi}\,H_{\alpha\mu\nu}
H^{\alpha\mu\nu} + \Lambda\,e^{2\phi} &= &0,\label{dilE}\\
d\,{}^{\star} H - 4\,d\phi\wedge{}^{\star} H &= &0.\label{axionE}
\end{eqnarray}
The right-hand side of Einstein's equations (\ref{einE}) is
represented by the energy-momentum tensor of the dilaton and axion
fields:
\begin{equation}
T_{\alpha\beta} =  2\,(D_\alpha\phi)(D_\beta\phi) - (D_\mu\phi)
(D^\mu\phi)\,g_{\alpha\beta} + e^{-4\phi}\left({\frac 1 4}
\,H_{\alpha\mu\nu}H_\beta{}^{\mu\nu} - {\frac 1
{24}}\,H_{\mu\nu\lambda}
H^{\mu\nu\lambda}\,g_{\alpha\beta}\right).\label{TEM}
\end{equation}
We will study the solutions of the effective string model in the
Einstein frame.

\section{Metric and the geometry of spacetime}

In this paper we continue the investigation of the cosmological
models belonging to the wide class of rotating spatially homogeneous
metrics \cite{rotrev}
\begin{equation}
ds^2 =  dt^2 - 2\,R\,n_{a}dx^{a}dt -
R^2\,\gamma_{ab}\,dx^{a}dx^{b}.\label{met0}
\end{equation}
Hereafter the indices $a,b,c =  1,2,3$ label the spatial coordinates,
$R =  R(t)$ is the scale factor, and
\begin{equation}
n_{a}= \nu_A\,e_a^{A},\qquad
\gamma_{ab}= \beta_{AB}\,e_a^{A}e_b^{B}.\label{ng}
\end{equation}
Here $\nu_{A},\beta_{AB}$ are constant coefficients ($A,B =  1,2,3$),
while
\begin{equation}
e^{A}=  e_{a}^{A}(x)\,dx^a \label{ea}
\end{equation}
are the invariant $1$--forms with respect to the action of a
three-parameter group of motion which is admitted by the space-time
(\ref{met0}). We assume that this group acts simply-transitively on
the spatial ($t= const$) hypersurfaces. It is well known that there
exist 9 types of such manifolds, distinguished by the Killing vectors
$\xi_{A}$ and their commutators
$[\xi_{A},\xi_{B}]= f^{C}{}_{AB}\,\xi_{C}$. The invariant forms
(\ref{ea}) solve the Lie equations ${\cal L}_{\xi_{B}} e^{A}= 0$ for
each Bianchi type, so that the models given by (\ref{met0}) are
spatially homogeneous.

The models of this class have rather attractive geometrical and
physical properties. In particular, we can recall that, in the
majority of the earlier cosmological models with rotation, shear
manifests itself in the parallax effects and in the distortion of the
background radiation which imposes very strong limits on the value of
the vorticity \cite{col}. In contrast, one can show \cite{rotrev}
that the parallax effects are absent and the microwave background
remains isotropic in all shear-free spacetimes (\ref{met0}). As a
result, these models satisfy all the known observational tests,
including the most important limits from the microwave background
radiation analysis \cite{col} which are not applicable to the
shear-free geometries. For further details on the observational tests
for such rotating models see \cite{coll} Moreover, these models are
completely causal (no closed timelike curves are allowed), when the
matrix $\beta_{AB}$ is positive definite.

Previously \cite{nine} we have studied Bianchi type IX closed {\it
stationary} worlds with the spinning fluid as the source. Now, we
consider {\it expanding closed worlds} belonging to the Bianchi type
IX. Denoting the spatial coordinates $x= x^1, y= x^2, z= x^3$, we have
explicitly the invariant 1-forms
\begin{eqnarray}
e^1 &= & \cos y\,\cos z\,dx - \sin z\,dy,\nonumber\\ e^2 &= & \cos
y\,\sin z\,dx + \cos z\,dy,\label{eA}\\ e^3 &= & -\,\sin y\,dx +
dz,\nonumber
\end{eqnarray}
which satisfy the structure equations
\begin{equation}
de^A =  f^A{}_{BC}\,e^B\wedge e^C,\quad {\rm with}\quad f^1{}_{23}
= f^2{}_{31} =  f^3{}_{12} =  1.
\end{equation}

After these preliminaries, we can write the ansatz for the line
element (\ref{met0})
\begin{equation}
ds^2 =  g_{\alpha\beta}\,\vartheta^\alpha\,\vartheta^\beta,
\qquad  g_{\alpha\beta} =  {\rm diag}(1, -1, -1, -1),\label{ds2}
\end{equation}
in terms of the orthonormal coframe 1-forms $\vartheta^\alpha$:
\begin{equation}
\vartheta^{\widehat{0}} =  dt - R\,\nu_A\,e^A,\
\vartheta^{\widehat{1}} =  R\,k_1\,e^1,\ \vartheta^{\widehat{2}} =
R\,k_2\,e^2,\ \vartheta^{\widehat{3}} =  R\,k_3\,e^3.\label{var}
\end{equation}
Here, $k_1, k_2, k_3$ are positive constant parameters. The Greek
indices $\alpha,\beta,\dots =  0,1,2,3$ hereafter label the objects
with respect to the orthonormal frame; the hats over indices denote
the separate frame components of these objects. Using (\ref{var})
in (\ref{ds2}), we find explicitly the $3\times 3$ matrix:
\begin{equation}\label{beta}
\beta_{AB} = \left(\begin{array}{ccc}k_1^2 - \nu_1^2 & -\nu_1\nu_2
& -\nu_1\nu_3\\ -\nu_1\nu_2 & k_2^2 - \nu_2^2 & -\nu_2\nu_3 \\
-\nu_1\nu_3 &-\nu_2\nu_3 & k_3^2 - \nu_3^2 \end{array}\right).
\end{equation}

It is straightforward to obtain the kinematical quantities which
describe the spacetime geometry. A direct calculation of the
vorticity, $\omega_{\mu\nu} =  h^{\alpha}{}_{\mu}h^{\beta}{}_{\nu}
\nabla_{[\alpha}u_{\beta]}$, shear,
$\sigma_{\mu\nu}= h^{\alpha}{}_{\mu}
h^{\beta}{}_{\nu}\nabla_{(\alpha}u_{\beta)} - {1\over 3}\,h_{\mu\nu}
\nabla_{\lambda}u^{\lambda}$, and the volume expansion $\theta=
\nabla_{\lambda}u^{\lambda}$, yields:
\begin{eqnarray}
&& \sigma_{\mu\nu} =  0,\qquad a^{\widehat{1}}= {\frac
{\dot{R}\nu_1}{Rk_1}}, \quad a^{\widehat{2}} =  {\frac
{\dot{R}\nu_2}{Rk_2}},\quad a^{\widehat{3}} =  {\frac
{\dot{R}\nu_3}{Rk_3}},\qquad \theta =  3{\frac {\dot{R}}{R}},\\ &&
\omega_{\widehat{2}\widehat{3}} =  -\,{\frac {\nu_1}{2Rk_2k_3}},\qquad
\omega_{\widehat{3}\widehat{1}} =  -\,{\frac {\nu_2}{2Rk_1k_3}},\qquad
\omega_{\widehat{1}\widehat{2}} =  -\,{\frac
{\nu_3}{2Rk_1k_2}}.\label{kin}
\end{eqnarray}
As usually, here $u= \partial_t$ is the velocity of a comoving
observer (normalized by $u_\alpha u^\alpha = 1$) and $h_{\mu\nu} =
g_{\mu\nu} - u_\mu u_\nu$ is the standard projector on his rest
3-space.

\section{Field equations and their exact solutions}

\subsection{Axion equation}

The axion field equation (\ref{axionE}) is solved by
\begin{equation}
H =  {\frac {a_0} {R^3}}\left(\vartheta^{\widehat{1}}\wedge
\vartheta^{\widehat{2}}\wedge\vartheta^{\widehat{3}} + {\frac
{\nu_1}{k_1}}\,\vartheta^{\widehat{0}}\wedge
\vartheta^{\widehat{2}}\wedge\vartheta^{\widehat{3}} + {\frac
{\nu_2}{k_2}}\,\vartheta^{\widehat{0}}\wedge
\vartheta^{\widehat{3}}\wedge\vartheta^{\widehat{1}} + {\frac
{\nu_3}{k_3}}\,\vartheta^{\widehat{0}}\wedge
\vartheta^{\widehat{1}}\wedge\vartheta^{\widehat{2}}\right).\label{Hfield}
\end{equation}
Note that one needs also to satisfy the condition $dH= 0$ which means
the existence of the potential 2-form so that $H= dB$. Here $a_0$ is
an arbitrary integration constant.

\subsection{Einstein-dilaton system}

The analysis of the Einstein-dilaton system (\ref{einE})-(\ref{dilE})
ultimately shows that the geometric parameters of the model should
satisfy
\begin{equation}
\nu_1\neq 0, \qquad \nu_2 =  \nu_3 =  0,\qquad k_2^2 =  k_3^2
=  k_1^2 - \nu_1^2.\label{kkk}
\end{equation}
The last equation is particularly important, as it guarantees the
absence of closed timelike curves since then the matrix (\ref{beta})
turns out to be explicitly positive definite: $\beta_{AB} =  k_2^2
\,\delta_{AB}$. Making use of (\ref{kkk}), one can verify that
(\ref{einE})-(\ref{dilE}) reduce to the system of the three equations
\begin{eqnarray}
{\frac {\ddot{R}}{R}} + 2\,{\frac {\dot{R}^2}{R^2}} + {\frac {k_1^2}
{2k_2^4\,R^2}}-{\frac {k_1^2}{k_2^2}}\,\Lambda\,e^{2\phi}
&= &0,\label{eq1}\\ -\,{\frac {\ddot{R}}{R}} + {\frac
{\dot{R}^2}{R^2}} + {\frac {k_1^2} {4k_2^4\,R^2}} -{\frac
{e^{-4\phi}a_0^2} {4R^6}}-\dot{\phi}^2 &= &0,\label{eq2}\\ \ddot{\phi}
+ 3\,{\frac {\dot{R}}{R}}\,\dot{\phi} - {\frac {e^{-4\phi} a_0^2}
{2R^6}} + {\frac {k_1^2}{k_2^2}}\,\Lambda\,e^{2\phi} &= &0.\label{eq3}
\end{eqnarray}
These three equations are not independent. There is a first integral
\begin{equation}
3{\frac {\dot{R}^2}{R^2}} + {\frac {3k_1^2}{4k_2^4\,R^2}} -
\dot{\phi}^2 - {\frac {k_1^2}{k_2^2}}\,\Lambda\,e^{2\phi} - {\frac
{e^{-4\phi} a_0^2} {4R^6}} =  0\label{eq4}
\end{equation}
which provides the consistency of the system (\ref{eq1})-(\ref{eq3}).
The unknown scale factor $R(t)$ and the dilaton field $\phi$ can be
determined from the integration of (\ref{eq4}) together with any one
of the equations (\ref{eq1})-(\ref{eq3}) (the two remaining equations
are then satisfied automatically).

\subsection{Simple solution}

Before we analyze the Einstein-dilaton system in detail, one can
readily find one particular solution:
\begin{equation} \label{SIMPLE}
e^\phi =  {\frac Q R},\qquad R =  R_0 + {\frac {k_1}{2k_2^2}}\,P\,t,
\qquad a_0 =  {\frac {k_1}{k_2^2}}\,Q^2.
\end{equation}
Here $Q$ and $P$ are the two arbitrary constants which have to
fulfill the condition
\begin{equation}
P^2 + 1 - 2\Lambda Q^2k_2^2 = 0.
\end{equation}
When $2\Lambda Q^2k_2^2 =  1$, the rotating world is stationary, but
for $2\Lambda Q^2k_2^2 > 1$ the cosmological scale factor
monotonously grows with the linear law.

\subsection{Convenient change of variables}

Let us introduce instead of the scale factor $R(t)$ and the dilaton
field $\phi(t)$ the new variables
\begin{equation}
\psi :=  \ln R + \phi,\qquad r:=  \ln R + {\frac 1
3}\,\phi.\label{change}
\end{equation}
If we simultaneously change the independent time variable from $t$ to
\begin{equation}
\tau :=  \int^t dt\,e^\phi,\qquad {\rm i.e.}\qquad {\frac d {d\tau}} =
e^{-\,\phi}\,{\frac d {dt}},
\end{equation}
then the system of the equations (\ref{eq4}) and (\ref{eq2}),
(\ref{eq3}) is re-casted into
\begin{eqnarray}
2\left({\frac {d\psi}{d\tau}}\right)^2 - 6\left({\frac
{dr}{d\tau}}\right)^2 + V(\psi) &= &0,\label{e1}\\ {\frac
{d^2\,r}{d\tau^2}} + \left({\frac {d\psi}{d\tau}}\right)^2
&= &0,\label{e2}\\ {\frac {d^2\,\psi}{d\tau^2}} + 3\,{\frac
{dr}{d\tau}}{\frac {d\psi}{d\tau}} + {\frac 1 4}\,{\frac {dV}{d\psi}}
&= &0,\label{e3}
\end{eqnarray}
where the potential $V= V(\psi)$ reads
\begin{equation}
V(\psi) =  {\frac {a_0^2} 3}\,e^{-6\,\psi} - {\frac {k_1^2}{k_2^4}}
\,e^{-2\,\psi} + {\frac {4k_1^2}{3k_2^2}}\,\Lambda.\label{V}
\end{equation}
The system (\ref{e1})-(\ref{e3}) has exactly the form of the
evolution equations in the so-called ``shifted frame'' introduced in
\cite{eas}. The potential (\ref{V}) has formally the same structure
as in \cite{eas}, but we have certain corrections of the coefficients
in $V$ by the rotation parameters.

In order to understand the behavior of the solutions, we need to
analyze the two subcases: (i) the central charge term dominating in
$V$ (this takes place for the large positive values of $\psi$), (ii)
the negligible (or zero) central charge term.

\subsection{Dominating cosmological term}

In the case when $\psi$ have large positive values, the potential
$V(\psi)$ is dominated by the central charge (cosmological) term and
the potential is approximately
$V= \frac{4}{3}\frac{k_1^2}{k_2^2}\Lambda$. The solution of
(\ref{e1})-(\ref{e3}) is straightforward and we obtain for the
positive $\Lambda$:
\begin{eqnarray}
\psi(\tau) &= & -\,{\frac 1
{\sqrt{3}}}\,\ln\,\tanh\left(A\tau/2\right) + \psi_0,\\ r(\tau) &= &
{\frac 1 3}\,\ln\,\sinh\left(A\tau\right) + r_0.
\end{eqnarray}
Here $\psi_0$ and $r_0$ are integration constants and
\begin{equation}
A =  {\frac {k_1}{k_2}}\,\sqrt{2\Lambda}.\label{AL}
\end{equation}
The original scale factor and the dilaton field then read:
\begin{eqnarray} \label{AL1}
R^2 &= & R_0^2\,\sinh\left(A\tau\right)\left(\tanh\left(A\tau/2\right)
\right)^{1/\sqrt{3}},\\  \label{AL2} e^{2\phi} &= & {\frac
{e^{2\phi_0}}{\sinh\left(A\tau\right)
\left(\tanh\left(A\tau/2\right)\right)^{\sqrt{3}}}}.
\end{eqnarray}
A solution with $\Lambda<0$ can be obtained with the help of the
evident analytical continuation. It is worthwhile to note that, since
the constant parameters are always given by (\ref{kkk}), the closed
timelike curves are absent for all values of the central charge.

The above formulas describe the behavior of the system near the
singularity ($R= 0$) when for small moments of $\tau$ the functions
$\psi$ and $r$ take very large values. In this regime, the
cosmological term is dominating in the potential $V$.

\subsection{Vanishing cosmological term}

Now let us analyze the case of the {\it critical string cosmology}
when $\Lambda = 0$. It is convenient to take (\ref{eq1}) and (\ref{eq4})
as the dynamical equations in this case. The equation (\ref{eq1}) then
can be straightforwardly integrated and yields the first integral:
\begin{equation}
R^4\left(\dot{R}^2 + {\frac {k_1^2}{4\,k_2^4}}\right) =  {\frac
{C_0^2} 3}.\label{C0}
\end{equation}
Here the integration constant is necessarily non-negative and we
denoted it by $C_0^2/3$ (the coefficient $1/3$ is introduced for
later convenience).

Introducing the dimensionless variables, $\rho:= R/R_0$ and
$\xi:= t/t_0$ with
\begin{equation}
R_0^4 =  \left({\frac {2C_0k_2^2}{\sqrt{3}k_1}}\right)^2,\qquad t_0^2
=  {\frac 3 {C_0^2}}\left({\frac {2C_0k_2^2}{\sqrt{3}k_1}}\right)^3,
\end{equation}
we reduce (\ref{C0}) to the quadratures
\begin{equation}
\int\,d\rho\,{\frac {\rho^2}{\sqrt{1 - \rho^4}}} = \,\pm\int\,d\xi.
\end{equation}
Integrating this, we obtain the solution $\rho(\xi)$ in the implicit
form:
\begin{equation} \label{VAN1}
E(\arcsin\rho\,|-\!1) - F(\arcsin\rho\,|-\!1) =  \,\pm(\xi - \xi_0),
\end{equation}
where $F(\varphi|m)$ and $E(\varphi|m)$ are (first and second kind)
elliptic integral special functions. Another (equivalent)
representation of the same solution uses the hypergeometric function:
\begin{equation} \label{VAN2}
{\frac {\rho^3} 3}\,{}_2F_1\!\left(\hbox{$\scriptsize{\frac 1 2}$},
\hbox{$\scriptsize{\frac 3 4}$}, \hbox{$\scriptsize{\frac 7 4}$};
\rho^4\right) =  \,\pm(\xi - \xi_0).
\end{equation}
The behavior of the scale factor as a function of time is depicted in
Fig.~1. 

\begin{figure}
\begin{center}
\resizebox{8 cm} {!}
   {\includegraphics{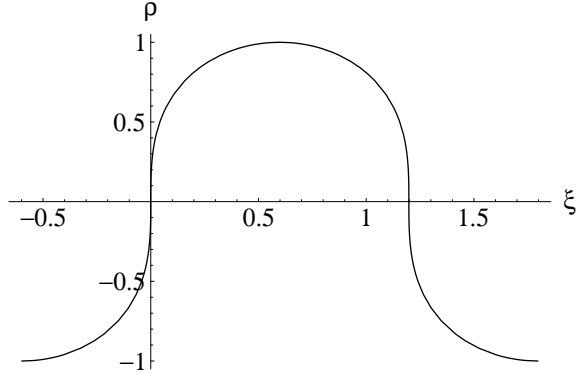}}
\caption{Scale parameter $\rho = R/R_0$ as a function of the
cosmological time $\xi = t/t_0$ (for $\xi_0= 0$).}
\end{center}
\end{figure}

It is now easy to find the dilaton field by making use of equation
(\ref{eq4}). Substituting (\ref{C0}) into (\ref{eq4}), we integrate
the resulting equation to find:
\begin{equation}
e^{2\phi}= {\frac 1{4C_0}}\left[\exp\left(2C_0\!\int{\frac
{dt}{R^3}}\right) + a_0^2\,\exp\left(-\,2C_0\!\int{\frac
{dt}{R^3}}\right)\right].
\end{equation}
One can do more and obtain the explicit form of the function
$e^{2\phi}(R)$ in terms of the scale factor. Using the above
solution, we derive:
\begin{equation} \label{VAN3}
e^{2\phi} =  {\frac 1 {4C_0}}\left[C_1\left({\frac {R^2} {1 + \sqrt{1
- (R/R_0)^4}}}\right)^{\pm\,\sqrt{3}} + {\frac {a_0^2}{C_1}}
\left({\frac {R^2}{1 + \sqrt{1 - (R/R_0)^4}}}\right)^{\mp\,\sqrt{3}}
\right].\label{dilR}
\end{equation}
Here $C_1$ is a new arbitrary integration constant.

Note that although the formulas obtained describe the exact solution
for $\Lambda = 0$, they also display the approximate behavior of the
system when the cosmological term can be neglected as compared to the
2 other terms in the potential $V$.

\subsection{Exact rotating solution}

The combination of the results of the two previous subsections yields
the complete dynamics of the axion-dilaton-graviton system for the
nontrivial $\Lambda$. However, besides the above approximate
configurations, there always exists an infinite family of the complete
{\it exact} solutions of (\ref{e1})-(\ref{e3}) which is parametrized
by an arbitrary real papameter. These solutions are obtained
when the cosmic rotation parameters satisfy the condition
\begin{equation} \label{FULL}
{\frac {k_1^2}{k_2^8}} =  8a_0^2\Lambda^2.\label{Lak}
\end{equation}
Then, denoting
\begin{equation}
e^{2\psi_0} :=  {\frac 1 {4k_2^2\Lambda}},
\end{equation}
we can introduce the ``Hubble function''
\begin{equation}
H =  \pm\,{\frac A 3}\left(1 - e^{2(\psi_0
-\psi)}\right)^{3/2}\label{H}
\end{equation}
[with the constant $A$ given by (\ref{AL})] such that the potential
(\ref{V}) is given by $V(\psi)= 6H^2-2(dH/d\psi)^2$. The system
(\ref{e1})-(\ref{e3}) is then evidently integrated by setting
$dr/d\tau= H$ and $d\psi/ d\tau =  -\,dH/d\psi$. As a result, the
solution can be given explicitly in the parametric form:
\begin{eqnarray}
r(\psi)-r(\psi_0)&= &-\int^{\psi}_{\psi_0}
\frac{H(\psi)}{H'(\psi)}d\psi,\label{e2t}\\ \tau(\psi)-\tau(\psi_0)
&= &-\int^{\psi}_{\psi_0}\frac{1}{H'(\psi)}d\psi.\label{e3t}
\end{eqnarray}
Using (\ref{H}), the integrals are straightforwardly evaluated to
give
\begin{eqnarray} \label{FULL1}
r(\psi)-r(\psi_0)&= &{\frac 1 3}(\psi-\psi_0)-{\frac 1 6}\left(
e^{2(\psi-\psi_0)} - 1\right),\\ \label{FULL2}
\tau(\psi)-\tau(\psi_0) &= & \mp {\frac 1 {2A}}\left[ e^{\psi-\psi_0}
\,\sqrt{e^{2(\psi - \psi_0)} -1} + \ln\left(e^{\psi-\psi_0} +
\sqrt{e^{2(\psi - \psi_0)} -1}\right)\right].
\end{eqnarray}
These results can be directly compared to the non-rotating solution
of \cite{eas} which is recovered in the limit of the vanishing
rotation parameter $\nu_1 =  0$. A qualitative new feature, as
compared to \cite{eas}, is that the value of the axion integration
constant $a_0$ is not rigidly fixed by the central charge $\Lambda$:
there always exists a nontrivial $\nu_1$ which satisfies (\ref{Lak})
for any $a_0$ and $\Lambda$.

\section{Discussion and conclusion}

In this paper, we have studied the problem of universal rotation in
string cosmology. We have demonstrated the existence of a number of
cosmological models with rotation {\it and} expansion which are exact
solutions of the low-energy effective string (non-critical, in
general) field equations for the metric, dilaton and axion. These
solutions are spatially homogeneous, belonging to the type IX of the
Bianchi classification. The models under consideration are
parallax-free and have an isotropic background radiation, being thus
in a good agreement with observations \cite{rotrev}. Moreover, the
solutions obtained are completely causal: the absence of closed
timelike curves is guaranteed by the positive definiteness of the
matrix $\beta_{AB} = k_2^2\,\delta_{AB}$ which follows from
(\ref{kkk}).

The magnitude of the universal rotation for all solutions is given by
\begin{equation}
\omega^2 =  {\frac{1}{2}}\omega_{\alpha\beta}\omega^{\alpha\beta} =
{\frac{\nu_1^2}{4R^2k_2^4}} =  \frac{1}{4R^2}\left(\frac{k_1^2}{k_2^4}
+ \frac{1}{k_2^2}\right).
\end{equation}
Correspondingly, the vorticity decreases with the growth of the scale
factor.

It is worthwhile to mention that, among other solutions, we derive
the simple solution (\ref{SIMPLE}) with the scale factor increasing
linearly in time. The (constant) expansion ``velocity'' $\dot{R}$ is
essentially determined by the rotation parameters, in particular, it
becomes faster with an increasing rotation.

Other solutions of the system (\ref{eq1})-(\ref{eq3}) turn out to be
closely related to the non-rotating solutions of \cite{eas} which is
based on the possibility to recast the original system into the form
(\ref{e1})-(\ref{e3}) of the so-called ``shifted frame'' introduced
in \cite{eas}. The analysis of the system (\ref{e1})-(\ref{e3}) shows
that it is sufficient to consider the two cases when $\Lambda$ is,
respectively, essentially larger and smaller than the other terms in
the ``potential'' (\ref{V}). In the case of a dominating central
charge (cosmological) term, the approximate behavior of the scale
factor and of the dilaton is given by (\ref{AL1}), (\ref{AL2}) which
describe the fields near the cosmological singularity at $R= 0$. As we
see, the large values of the rotation parameter $\nu_1$ yield a more
rapid growth of the scale factor, whereas the vorticity quickly
decreases in time.

When the cosmological term vanishes, we derive an exact solution for
the system (\ref{eq1})-(\ref{eq3}). It is given by the equations
(\ref{VAN1}) or (\ref{VAN2}) and (\ref{VAN3}). For a large value of
rotation constants ($k_1\gg k_2$), we again see that the scale factor
increases rapidly in time, whereas the vorticity quickly dilutes
during the expansion. Such a behavior is also approximately correct
for small values of the cosmological term in (\ref{V}).

Finally, there always exists an exact solution with cosmic rotation
parameters fulfilling the condition (\ref{FULL}). It is possible to
find a suitable $\nu_1$ for any values of the axion constant $a_0$
and the cosmological term $\Lambda$. These exact solutions are given
by (\ref{FULL1}), (\ref{FULL2}) and they represent the direct
generalization of the non-rotating solutions discussed earlier in
\cite{eas}.

Summarizing, we find that all the exact solutions which describe the
expanding and rotating cosmological models have the same qualitative
properties: The large rotation parameters accelerate the expansion of
the world, or, in the case of a contracting phase, this decelerates
the contraction rate. However, the rotation apparently cannot prevent
the occurrence of cosmological singularities, at least in the low
energy approximation. The study of higher order corrections to the
effective string action may change this conclusion.

The last but not least remark: Since the solutions obtained describe
the spatially closed world, their very existence represents the
evidence of the possibility of the truly anti-Machian cosmological
models. This observation extends the previous results
\cite{osch,infl,nine} to the case of the more physically realistic
closed models with rotation {\it and} expansion.

\section{Acknowledgments}

The authors would like to thank A. Tseytlin for the useful hints and
comments about the effective string theory. This work was supported
by the Deutsche Forschungsgemeinschaft with the grant 436 RUS
17/70/01.


\end{document}